\documentstyle[aps,epsfig,tighten,twoside,psfig]{revtex}

\input epsf

\newcommand{\be}{\begin{equation}}
\newcommand{\en}{\end{equation}}
\newcommand{\bea}{\begin{eqnarray}}
\newcommand{\ena}{\end{eqnarray}}

\newcommand{\hbo}{\hbox to 1 true cm {\hfill } }
\newcommand{\tr}{\hbox{tr}}

\begin{document}

\vglue 1truecm
  
\vbox{ UNITU-THEP-009/2001 
\hfill March 29, 2001
}
  
\vfil
\centerline{\large\bf SU(2) gluon propagators from the lattice -- a preview } 
  
\bigskip
\centerline{ Kurt Langfeld } 
\vspace{1 true cm} 
\centerline{ Institut f\"ur Theoretische Physik, Universit\"at 
   T\"ubingen }
\centerline{D--72076 T\"ubingen, Germany}
  
\vfil
\begin{abstract}
High accuracy numerical results for the SU(2) gluonic form factor 
are previewed for the case of Landau gauge. 
I focus on the information of quark confinement encoded in the 
gluon propagator.

\end{abstract}

\vfil
\hrule width 5truecm
\vskip .2truecm
\begin{quote} 


keywords: {\it Landau gauge, Gribov problem, gluon propagator, confinement, 
lattice gauge theory. }

\end{quote}
\eject

{\bf Introduction.}
Two prominent methods to treat non-perturbative Yang-Mills theory will be 
addressed in this talk: the numerical simulation of lattice gauge theory 
(LGT) and the approach by means of the Dyson-Schwinger equations (DSE). 
While LGT covers all non-perturbative effects and, in particular, bears 
witness of quark confinement (see e.g.~\cite{bali95}), 
simulations including dynamical quarks are cumbersome despite the 
recent successes by improved algorithms~\cite{kap92} and the 
increase of computational power. 
At the present stage, systems at finite baryon densities are hardly 
accessible in the realistic case of an SU(3) gauge group~\cite{bar98}
(for recent successes see~\cite{eng99}). By contrast, the DSE approach 
can be easily extended for an investigation of quark 
physics~\cite{rob94,alk00} 
even at finite baryon densities~\cite{rob00}. Unfortunately,
the DSE approach requires a truncation of the infinite 
tower of equations, and this approximation is difficult to improve 
systematically. 
In addition, the DSE approach needs gauge fixing which is obscured by 
Gribov copies. Whether the standard Faddeev-Popov method of gauge fixing 
is appropriate in non-perturbative studies, is still under 
debate~\cite{bau98}. 

\vskip 0.3cm 
In order to merge the advantages of both approaches to low energy 
Yang-Mills theory, I will firstly address the gluon propagator 
of pure SU(2) lattice gauge theory in Landau gauge. The result 
can then be compared with the one provided by the solution of the 
coupled ghost-gluon Dyson equation~\cite{sme97,atk98,atk98b}. This will allow 
us to estimate the soundness of the truncations introduced 
to solve the equations (e.g.~vertex ansatz, angular approximation). 
Secondly, the gluon propagator is an 
one essential ingredient of the quark DSE. Two options are obvious: 
a parameterization of the lattice result for the gluon propagator 
enters the quark DSE. The corresponding solution of this equation 
provides informations on hadronic observables in quenched approximation 
i.e.~the backreaction of the quarks on the gluonic Greenfunctions 
is neglected. Once the reliability of the DSE approach to the ghost 
gluon system has been tested, the second option is to solve 
a truncated set of coupled ghost-gluon-quark DSEs, thereby, challenging 
the quenched approximation. 

\vskip 0.3cm 
In my talk, I will focus on the gluon propagator as inferred from 
the lattice calculation, and I will concentrate on the information on 
quark confinement which might be encoded in the gluon propagator. 
High accuracy data for the latter 
are obtained by a numerical method superior to 
existing techniques. Further informations and details of the numerical 
method will be presented in a forthcoming publication.

\vskip 0.3cm 
{\bf Lattice definition of the gluon field.} 
Before identifying the gluonic degrees of freedom in the lattice 
formulation, I briefly recall the definition of the gluon field 
in continuum Yang-Mills theory. 

\vskip 0.3cm 
The starting point for constructing Yang-Mills theories is the 
transformation property of the matter fields. In the case of an 
SU(2) gauge theory, we demand invariance under local SU(2) (say color) 
transformations of the quark fields 
\be 
q^\prime (x) \; = \; \Omega \, (x) \; q(x) \; , \hbo 
\Omega (x) \in {\mathrm SU(2) } \; . 
\label{eq:1} 
\en 
In order to construct a gauge invariant kinetic term for the 
quark fields, one defines the gauge covariant derivative $D_\mu 
:= \partial _\mu + i A_\mu^a t^a$, where $t^a$ are the generators 
of the SU(2) gauge group. Per definitionem, this covariant 
derivative homogeneously transforms under gauge transformations, 
\be 
D^\prime _\mu q^\prime (x) \; = \; \Omega (x) \, D_\mu (x) \, q(x) 
\label{eq:2} 
\en 
if the gluon fields transforms as 
\bea 
A^{a \, \prime }_{\mu }(x) &=& O^{ab}(x) \,  A^b _\mu (x) 
\; + \; f^{aef} \, O^{ec}(x) \, \partial _\mu O^{fc} \; , 
\label{eq:3a} \\ 
O^{ab}(x) &:=& 2 \, \tr \bigl\{ \Omega (x) \, t^a \, 
\Omega ^\dagger (x) \, t^b \bigr\} \; , \hbo 
O^{ab}(x) \, \in \, {\mathrm SO(3) } \; . 
\label{eq:3} 
\ena 
The crucial observation is that the gluon fields transform according 
to the {\it adjoint} representation while the matter fields are 
defined in the fundamental representation. 

\vskip 0.3cm 
Let us compare these definitions of fields with the ones in LGT. 
In LGT, a discretization of space-time with a lattice spacing $a$ 
is instrumental. The 'actors' of the theory are SU(2) matrices 
$U_\mu (x)$ which are associated with the links of the lattice. 
These links transform under gauge transformations as 
\be 
U^\prime _\mu (x) \; = \; \Omega (x) \, U_\mu (x) \, \Omega ^\dagger  
(x+\mu)  \; \hbo \Omega (x) \in {\mathrm SU(2) } \; . 
\label{eq:4} 
\en 

For comparison with the ab initio continuum formulation, I also 
introduce the adjoint links 
\bea 
\widetilde{U}_{\mu }^{ab} (x) &:=&  2 \, \tr \bigl\{ U_\mu (x) \, t^a \, 
U^\dagger _\mu (x) \, t^b \bigr\} \; , 
\label{eq:5} \\ 
\widetilde{U}_{ \mu }^{\prime }(x) &:=&  O(x) \, \widetilde{U}_{\mu } (x) \, 
O^T(x) \; , \hbo O(x) \in {\mathrm SO(3) } \; , 
\ena 
where $O(x)$ was defined in (\ref{eq:3}). 

\vskip 0.3cm 
In order to define the gluonic fields from lattice configurations, 
I exploit the behavior of the (continuum) gluon fields under 
gauge transformations (see~(\ref{eq:3a})), and identify the lattice 
gluon fields $\hat{A}_\mu ^a (x)$ as algebra fields of the adjoint 
representation, i.e. 
\be 
\widetilde{U}_{\mu }^{ab} (x) \; =: \; \biggl[ \exp \bigl\{ \epsilon ^f  
\, \hat{A}^f_\mu (x) \, a \bigr\} \, \biggr] ^{ab} \; , 
\hbo \big(\epsilon ^f \bigr)^{ac} := \epsilon ^{afc} \; , 
\label{eq:6} 
\en 
where the total anti-symmetric tensor $\epsilon ^{abc}$ acts as generator 
for the SU(2) adjoint representation, and where $a$ denotes the lattice 
spacing.

\vskip 0.3cm 
For later use, it is convenient to have an explicit formula for the 
(lattice) gluon fields $\hat{A}_\mu ^a (x)$ in terms of the 
SU(2) link variables $U_\mu (x)$. Usually, these links are given  in terms 
of four vectors of unit length 
\be 
U_\mu (x) \; = \; u_\mu ^0(x) \, + \, i \, \vec{u}_\mu (x) \, 
\vec{\tau} \; , \hbo \bigl[ u^0_\mu (x)\bigr]^2 \, + \, 
\bigl[ \vec{u}_\mu (x)\bigr]^2 \; = \; 1 \; , 
\label{eq:7} 
\en 
where $\tau ^b$ are the Pauli matrices. 
Using these variables, a straightforward calculation yields 
\be 
\hat{A}^b_\mu (x) \, a \; + \; {\cal O}(a^2) \; = \; u^0_\mu (x) 
\, u^b_\mu (x) \; , \hbox to 4cm{ without summation over  } \, \mu \; .
\label{eq:8} 
\en 
I point out that (\ref{eq:8}) is a novel definition of the lattice 
gluon field. 

\vskip 0.3cm 
Finally, I illustrate the definition (\ref{eq:8}). Let us assume that 
we have exploited the gauge degrees of freedom (see (\ref{eq:4})) 
to bring the SU(2) link elements $U_\mu (x)$ as close as possible to 
the unit matrix, 
\be 
\Omega (x): \; \sum _{\{x\}, \mu } \tr \, U^\prime _\mu (x) 
\rightarrow \, \mathrm{max} \; .
\label{eq:9} 
\en 
In this gauge, I decompose the link variables by 
\be 
U^\prime _\mu (x) \; = \; Z_\mu (x) \, \exp \biggl\{ i \hat{A}^b_\mu (x) 
\, t^b \, a \biggr\} \; , 
\label{eq:10} 
\en 
where $\hat{A}^b_\mu (x) $ is implicitly defined by (\ref{eq:6}) and 
$Z_\mu (x) \in \{-1,+1\}$. Indeed, the lattice gluon fields (\ref{eq:8}) 
do not change when $U_\mu (x) \rightarrow (-1) U_\mu(x)$. Hence, the 
fields $\hat{A}^b_\mu (x) $ are relegated to the SO(3) coset space. 
I here propose 
to disentangle the information carried by the center and coset 
parts of the link variables by studying the $Z_\mu(x)$ and $\hat{A}^b_\mu (x)$ 
correlation functions independently. I stress, however, that in Landau 
gauge (\ref{eq:9}) the role of the $Z_\mu (x)$ is de-emphasized 
($Z_\mu (x) \rightarrow 1$). 
In particular, I do not expect a vastly different gluon propagator 
when other (more standard) definitions of the lattice gluon fields 
are used~\cite{cuc97,bon01}. 

\vskip 0.3cm 
{\bf Gauge fixing.} 
In the continuum formulation, calculations employing gauge fixed 
Yang-Mills theory use only gauged gluon fields which satisfies the 
gauge condition, e.g. 
\be 
\partial _\mu A^{\prime \, a }_\mu (x) \; = 0 \; \hbox to 3cm 
{\hfill (Landau gauge) \hfill } \; , 
\label{eq:11} 
\en 
and rely on the assumption that the Faddeev-Popov determinant corrects 
the probabilistic weight in an appropriate way. This assumption is true 
if the gauge condition picks a unique solution $\Omega (x)$ of (\ref{eq:11}) 
for a given field $A^a_\mu (x)$. Unfortunately, the Landau gauge 
condition generically admits several solutions depending on the 
''background field'' $A_\mu^b(x)$ which is the subject of gauge fixing 
(Gribov problem). Thus, the above assumption 
seems not always be justified~\cite{bau98}. Further restrictions 
on the space of possible solutions are required~\cite{zwa94}.

\vskip 0.3cm 
Let us contrast the continuum gauge fixing with its lattice analog. 
In a first step, link configurations $U_\mu(x)$ are generated by means 
of the gauge invariant action without any bias to a gauge condition. 
In a second step, the gauge-fixed ensemble is obtained by 
adjusting the gauge matrices $\Omega (x)$ (see~(\ref{eq:4})) 
until the gauged link ensemble satisfies the gauge condition. 
This procedure obviously guarantees the correct probabilistic measure 
for the gauged configurations, and gauge invariant quantities 
which are calculated with the gauged configurations evidently 
coincide with the ones obtained from un-fixed configurations. 
However, the Gribov problem re-appears as the problem of finding ''the 
name of the gauge''. Let me illustrate this last point: 
The naive Landau gauge condition for the lattice gluon field, i.e. 
\be 
\hat{\partial }_\mu \hat{A}^{\prime \, b} _\mu (x) \; = \; 0 
\label{eq:12} 
\en 
is satisfied if we seek an {\it extremum} of the variational condition 
(\ref{eq:9}). If we restrict the variety of solutions $\Omega (x)$ 
which extremize (\ref{eq:9}) to those solutions which {\it maximize } 
the functional (\ref{eq:9}), we confine ourselves to the case where 
the Faddeev-Popov matrix is positive semi-definite. The fraction of 
the configuration space of gauge fixed fields $\hat{A}^{\prime \, b} _\mu $ 
is said to lie within the first Gribov horizon. A numerical algorithm 
which obtains the gauge transformation matrices $\Omega (x)$ from 
the condition (\ref{eq:9}) still samples a particular set of local 
maxima. Different algorithms might differ in the subset of chosen 
local maxima, and, hence, implement different gauges. A conceptual 
solution to the problem is to restrict the configuration space 
of gauge fixed fields $\hat{A}^{\prime \, b} _\mu $ to the 
so-called {\it fundamental modular region}. In the lattice simulation, 
this amounts to picking the {\it global maximum } of the 
variational condition (\ref{eq:9}). In practice, finding the 
global maximum is a highly non-trivial task. Here, I adopt two extreme 
cases of gauge fixing: firstly, I will study the gluon propagator of 
the gauge where an iteration over-relaxation algorithm almost 
randomly averages over the local maxima of the variational condition 
(\ref{eq:9}). This result will then be compared with the gluon propagator 
of a gauge where a simulated annealing algorithm searches for the 
global maximum. It will turn out that the gluon propagators of both 
gauges agree within statistical error bars. 

\vskip 0.3cm
{\bf The numerical simulation:} 
The link configurations are generated using the Wilson action. 
I refrain from using a ''perfect action'' since I am interested in 
the gluon propagator in the full momentum range; simulations 
using perfect actions recover a good deal of continuum physics 
at finite values of the lattice spacing at the cost of a non-local action. 
For practical simulations, perfect actions are truncated which is 
poor approximation at high energies where the full non-locality of 
the action must come into play. 

\vskip 0.3cm
Here, calculations were performed using a $16^3 \times 32$ lattice. 
The dependence 
of the lattice spacing on $\beta $ (renormalization), i.e. 
\be 
\sigma a^2 (\beta ) \; = \; 0.12 \, \exp \biggl\{ - \frac{ 6 \pi^2}{11} 
\, \bigl( \beta - 2.3 \bigr) \biggl \} \, , \hbo 
\sigma := (440 \,{\mathrm MeV}) ^2 \; , 
\label{eq:20} 
\en 
is appropriate for $\beta \in [2.1, 2.6]$ for the achieved numerical
accuracy. 

\vskip 0.3cm
Once gauge-fixed ensembles are obtained by implementing a variational 
gauge condition (see discussion of previous section), the gluon 
propagator is calculated using 
\be 
D^{ab}_{\mu \nu } (x-y) \; = \; \bigl\langle \hat{A}^a_\mu (x) \, 
\hat{A}^b_\nu (y) \, \bigr\rangle _{MC} \; , 
\label{eq:21} 
\en 
where the Monte-Carlo average is taken over 200 properly thermalized 
gauge configurations. Of particular interest is the gluonic form factor 
$F(p^2)$ which is defined by 
\be 
D(p^2) \; = \; \int D^{aa}_{\mu \mu } (x) \; \exp \bigl\{ ipx \bigr\} 
\; d^4x \; , \hbo 
D(p^2) = \frac{F(p^2)}{p^2} \; . 
\label{eq:22} 
\en 
Since in Landau gauge the propagator is diagonal in color and transversal 
in Lorentz space, the form factor $F(p^2)$ contains the full information. 
\begin{table}
\caption{ Simulation parameters: $L=32 a$: lattice extension, 
$\Lambda $: UV cutoff } 
\begin{tabular}{cccccc}
$\beta $ & 2.1 & 2.2 & 2.3 & 2.4 & 2.5  \\ 
L [fm] & 8.6 & 6.6 & 5.0 & 3.8 & 2.9 \\ 
$\Lambda $ [GeV] & 2.3 & 3.0 & 4.0 & 5.2 & 6.8 \\
\end{tabular}
\end{table}

\vskip 0.3cm 
{\bf Results I: Landau gauge } 
The lattice momentum in units of the lattice spacing is given by 
\be 
p_x \, a \; = \; 2 \, \sin \left( \frac{\pi}{N} n_x \right) \; , 
\hbo x = 1 \ldots 4 
\label{eq:23} 
\en 
where $n_x$ is an integer which numbers the Matsubara frequency and 
$N$ is the number of lattice points (in $x$-direction). 
\begin{figure}[t]
\centerline{
\epsfxsize=0.5\linewidth 
\epsfbox{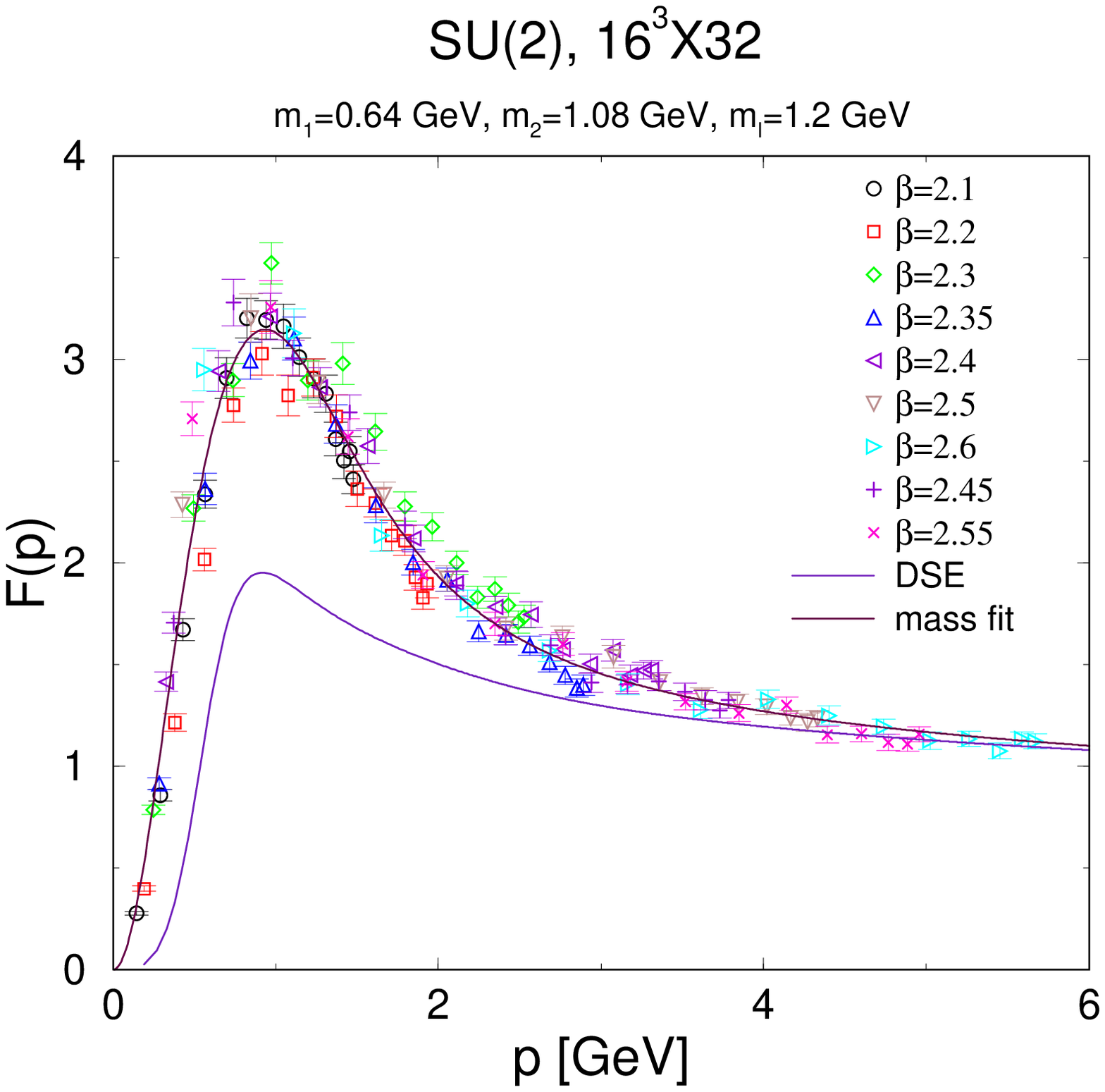}
\epsfxsize=0.55\linewidth 
\epsfbox{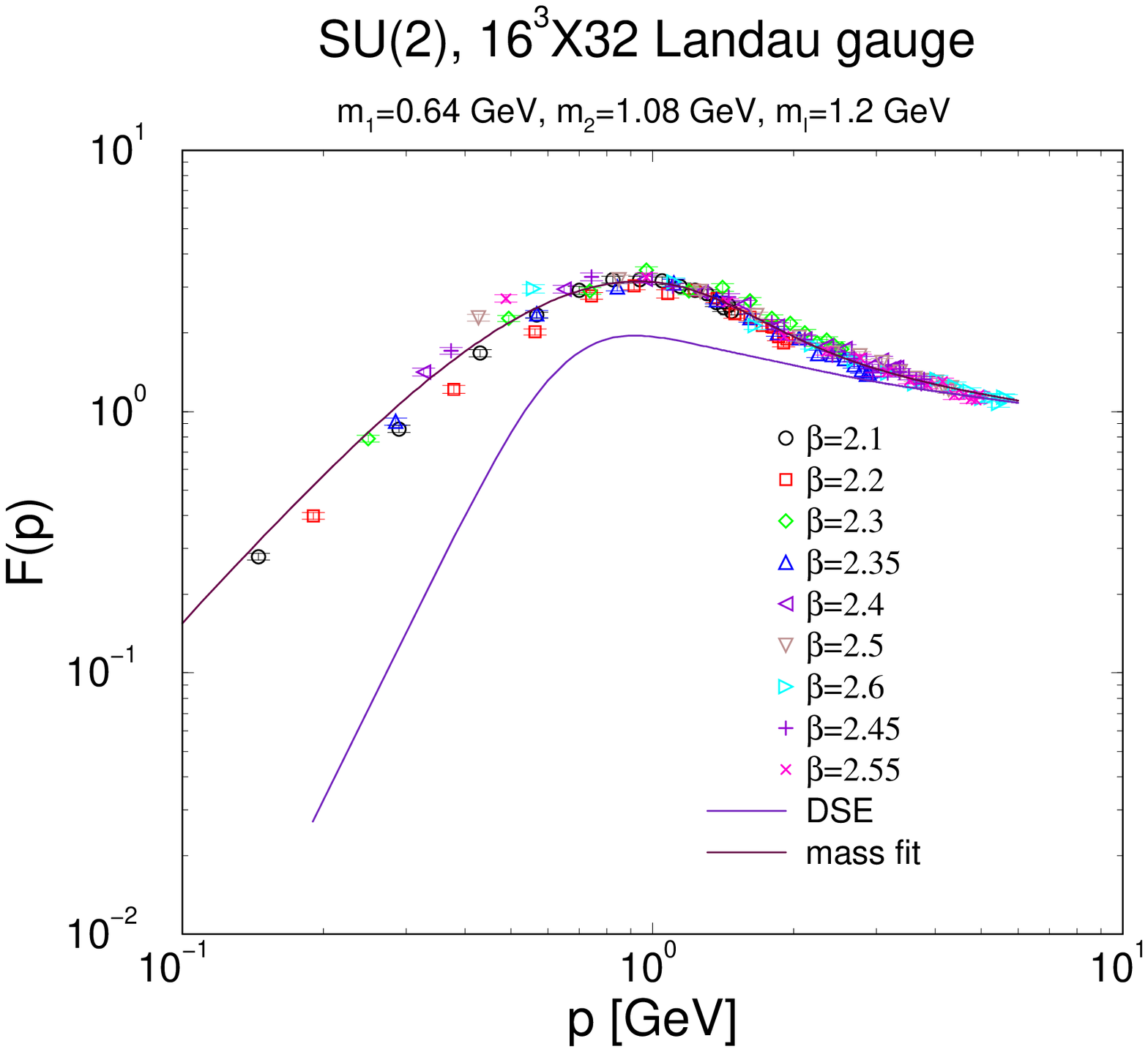}
}
\caption{The gluonic form factor $F(p^2)$ as function of the momentum 
   transfer (left panel: linear scale; right panel: log-log scale). 
   Also shown is the solution of the set of DSEs proposed 
   in [9] which have been solved for the case of $SU(2)$ [10].
}
\label{fig:1}
\end{figure}
Physical 
units for the momentum can be obtained by using (\ref{eq:20}). 
Calculations 
with different $\beta $-values correspond to simulations with a  
different UV-cutoff $\Lambda := \pi / a(\beta )$. In order to 
obtain the {\it renormalized} gluon propagator, the gluonic wave 
function renormalization is chosen to yield a finite (given) value 
for the form factor at fixed momentum transfer. 

\vskip 0.3cm
Figure \ref{fig:1} shows my result for the form factor $F(p^2)$ where 
the condition (\ref{eq:9}) was implemented with an iteration over-relaxation 
method. The data obtained with different $\beta $-values are identical 
within numerical accuracy, thus signaling a proper extrapolation 
to the continuum limit.

At high momentum the lattice data are consistent with the behavior 
known from perturbation theory, 
\be 
F(p^2) \; \propto \; 1/ \left[ log \frac{p^2}{\mu ^2 } \right]^{13/22} \; , 
\hbo p^2 \gg \mu ^2 \approx (1 \, \mathrm{GeV})^2 \; . 
\label{eq:24} 
\en 
The lattice data are compared with the solution of the gluon-ghost 
DSE~\cite{sme97}\footnote{ I thank Chr.~Fischer for communicating his 
DSE solution for the SU(2) case prior to publication.}. 
From the DSE studies one expects a scaling law at small momentum
\be 
F(p^2) \; \propto \left[ p^2 \right]^{2 \kappa } \; , \hbo 
p^2 \ll \mu ^2 \approx (1 \, \mathrm{GeV})^2 \; . 
\label{eq:25}  
\en 
Depending on the truncation of the Dyson tower of equations and on 
the angular approximation of the momentum loop integral, one finds 
$\kappa = 0.92$~\cite{sme97} or $\kappa = 0.77$~\cite{atk98} 
or $\kappa \rightarrow 1$~\cite{atk98b}. The 
lattice data are consistent with $\kappa = 0.5$ corresponding to 
an infrared screening by a gluonic mass. Also shown is the coarse grained 
''mass fit'' ($\mu = 1\, $GeV) 
\be 
F(p^2) \; = \; \frac{ p^2}{p^2+m_1^2} \biggl[ \frac{ \mu ^4 }{p^4+m_2^4} 
\; + \; \frac{s}{\left[ log \left(\frac{m_l^2}{\mu ^2 } + \frac{p^2}{\mu ^2 } 
\right) \right]^{13/22} } \biggr] 
\label{eq:26} 
\en 
which nicely reproduces the lattice data within the statistical error bars. 

\vskip 0.3cm 
{\bf Results II: gluon propagator and confinement } 
In order to get a handle on the information of quark confinement 
encoded in the gluon propagator in Landau gauge, I now {\it change by hand} 
the SU(2) Yang-Mills theory to a theory which does not confine quarks. 
It is instructive to compare the gluon propagator of the modified 
theory with the SU(2) result (see figure \ref{fig:1}). 

\vskip 0.3cm 
In the Maximal Center gauge~\cite{deb97}, the mechanism of quark confinement 
can be understood by a percolation of vortices which acquire physical 
relevance in the continuum limit~\cite{la98}. An intuitive picture 
in terms of vortex physics is also available for the deconfinement phase 
transition at finite temperatures~\cite{la99}. Reducing the full Yang-Mills 
configurations to their vortex content still yields the full string 
tension~\cite{deb97}. Vice versa, removing these vortices from the 
Yang-Mills ensemble results in a vanishing string tension. 
This observation was used in~\cite{for99} to verify the impact of the 
vortices on chiral symmetry breaking. 

\begin{figure}[t]
\centerline{
\epsfxsize=0.53\linewidth 
\epsfbox{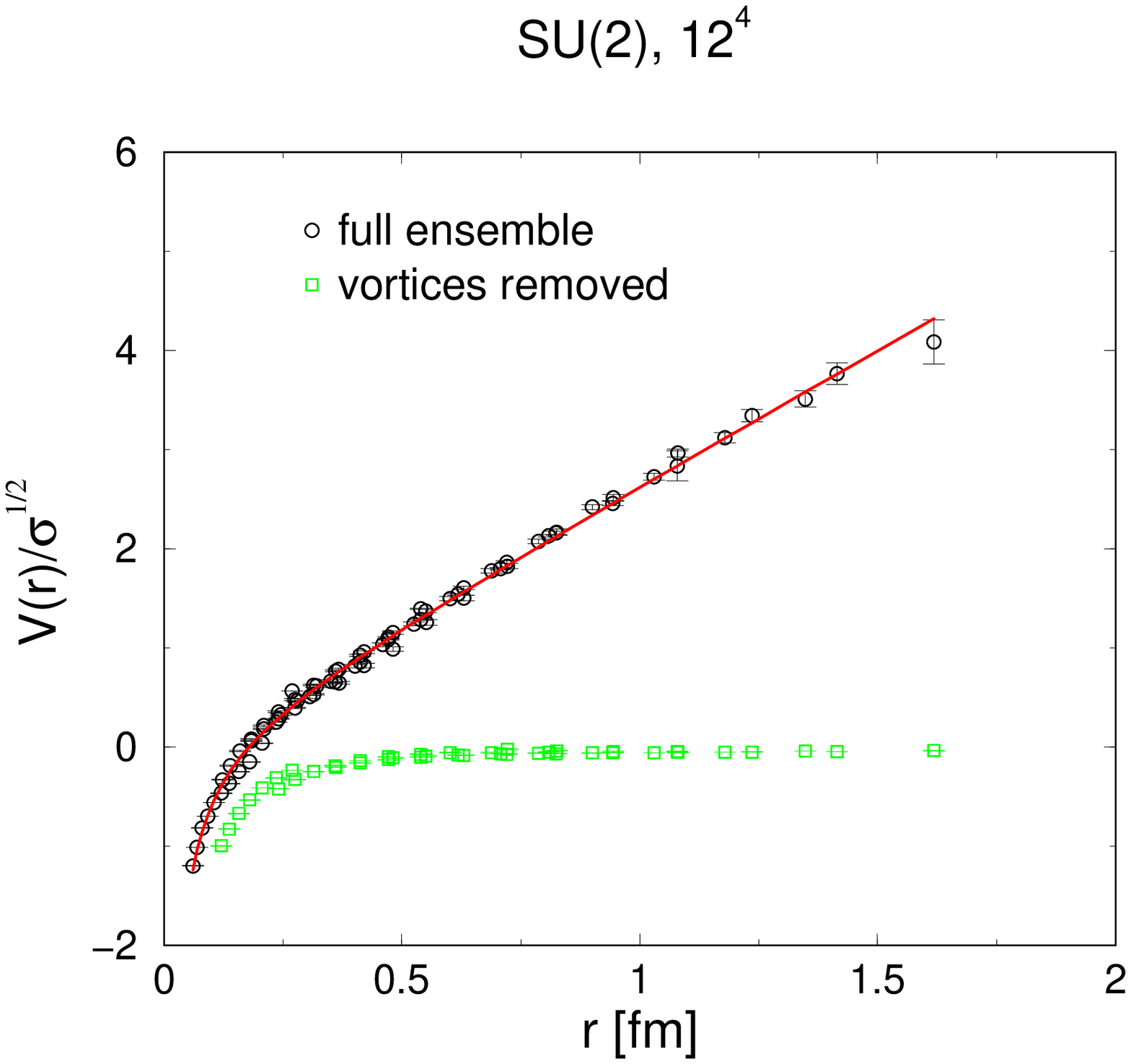}
\epsfxsize=0.5\linewidth 
\epsfbox{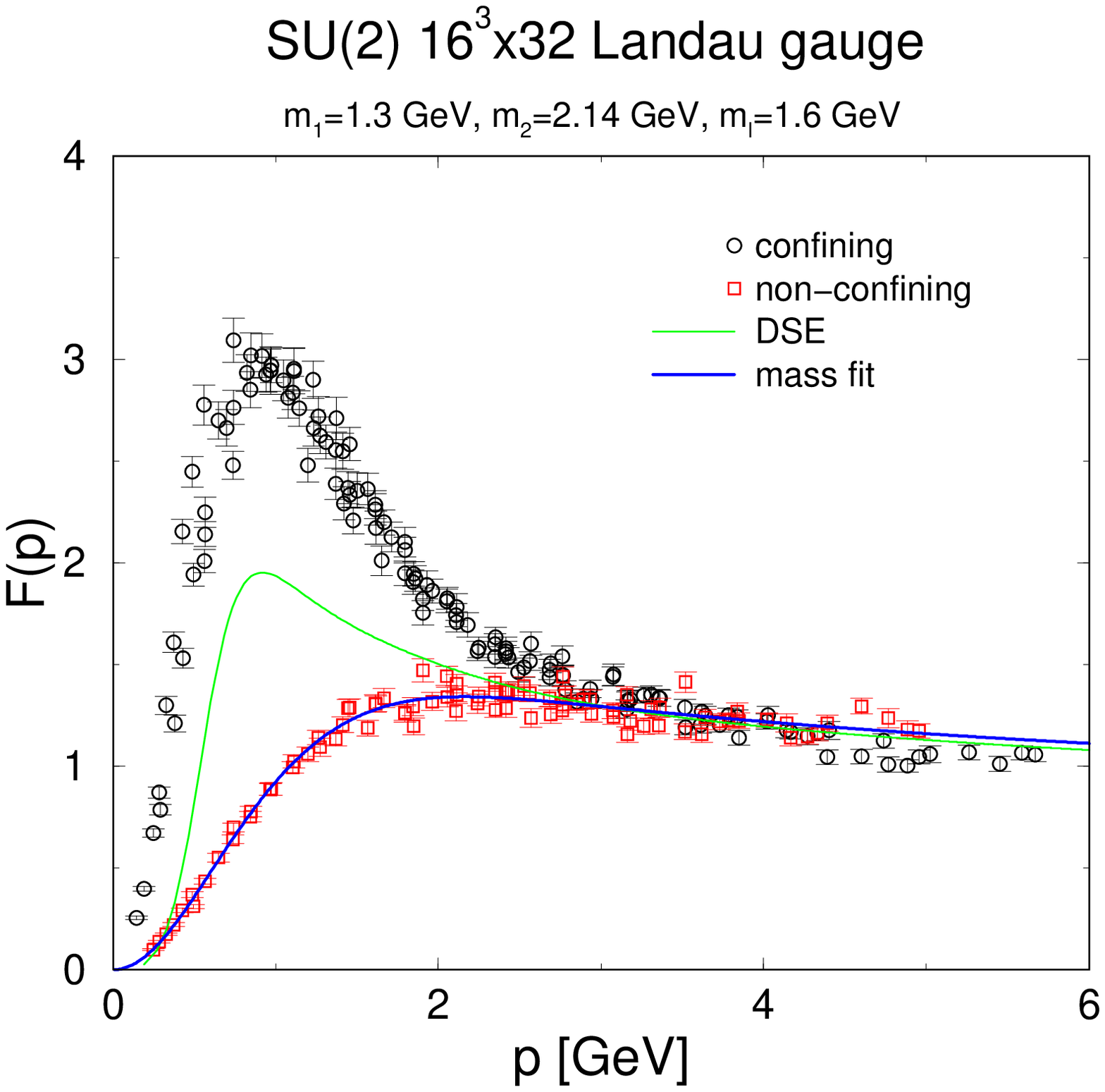}
}
\caption{ The static quark anti-quark potential (left panel) and 
   the corresponding gluonic form factors (right panel); DSE solution 
   from [10]. } 
\label{fig:2}
\end{figure}
\vskip 0.3cm 
The static quark anti-quark potential in figure \ref{fig:2} demonstrates 
that a removal of the center vortices produces a 
non-confining theory. Figure \ref{fig:2} also shows the gluonic form 
factor obtained from the modified ensemble. The striking feature is 
that the strength of the form factor in the infra-red momentum range 
is drastically reduced. 

\begin{figure}[t]
\centerline{
\epsfxsize=0.5\linewidth 
\epsfbox{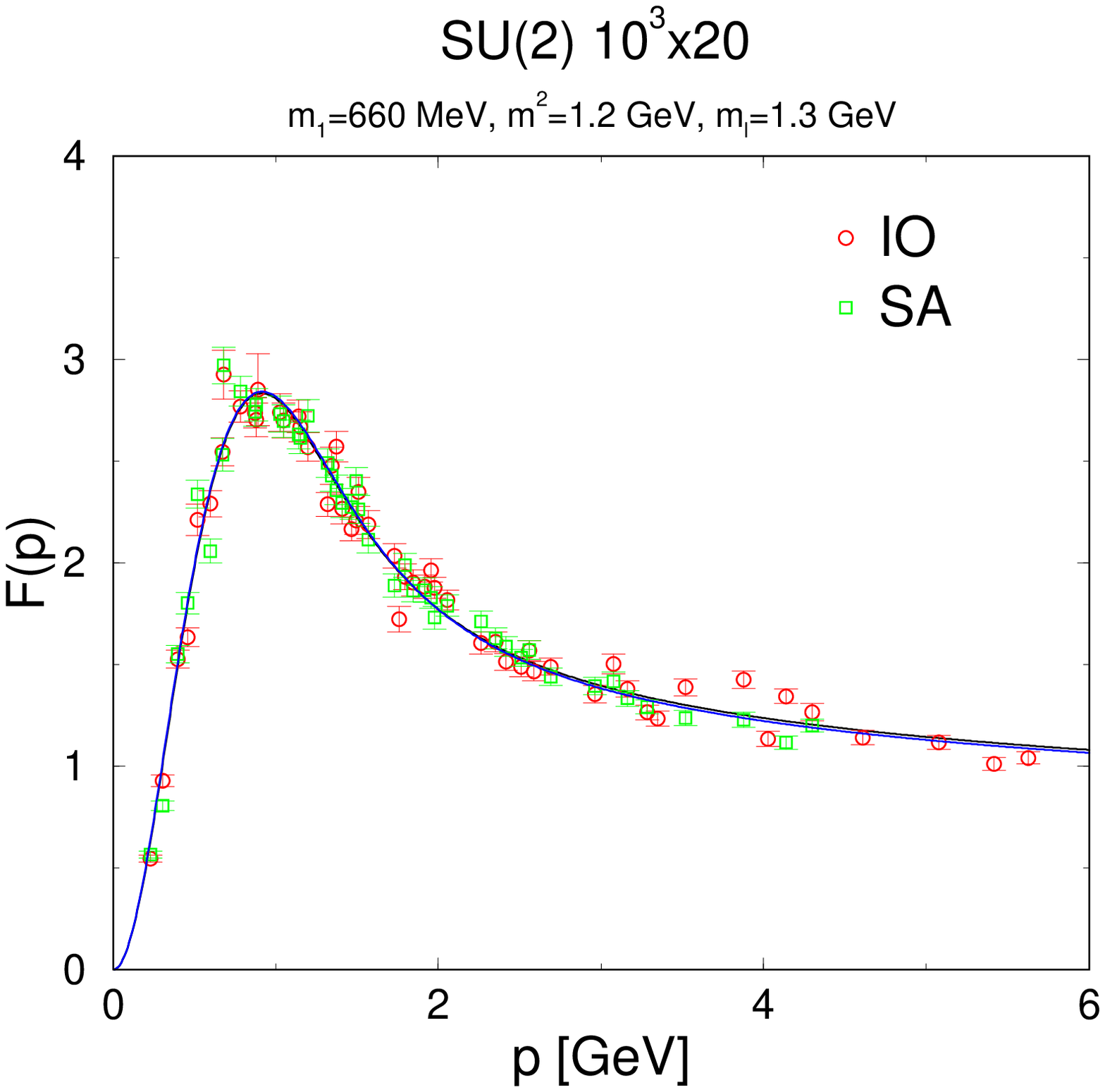}
\epsfxsize=0.5\linewidth 
\epsfbox{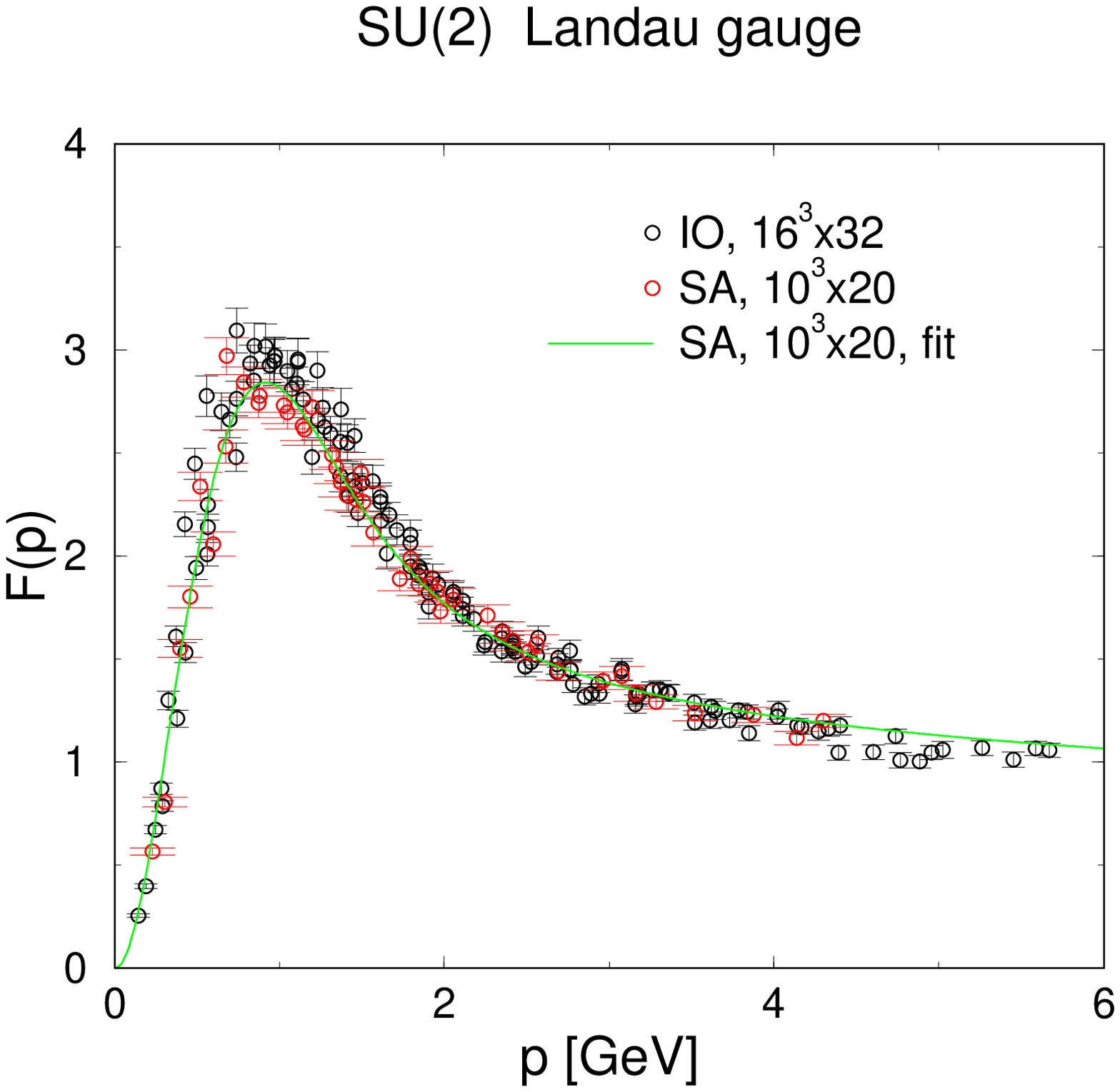}
}
\caption{The gluonic form factor $F(p^2)$ for a $10^3\times 20$ lattice 
   in the gauge IA and SA, respectively (left panel) and compared with 
   previous results (IA, $16^3\times 32$) (right panel). } 
\label{fig:3}
\end{figure}

\vskip 0.3cm 
{\bf Results III: the Gribov noise } 
Finally, let us check how strongly the gluonic form factor $F(p^2)$ 
depends on the choice of gauge, i.e. on the sample of maxima of the 
variational condition (\ref{eq:9}) selected by the algorithm. 
For this purpose, I adopt an extreme point of view by comparing the gauge 
implemented by the iteration over-relaxation (IA) algorithm with the 
gauge obtained by simulated annealing (SA). The results of the form factor 
in both cases are shown in figure~\ref{fig:3}. I find, in agreement 
with~\cite{cuc97}, that, in the case of the gluonic form factor, 
the Gribov noise is comparable with statistical noise for data generated 
with $\beta \in [2.1,2.5]$ (scaling window). 

\vspace{1cm}
{\bf Acknowledgements.} 
I thank my coworkers J.~Gattnar and H.~Reinhardt. I am indebted to 
J.~Bloch, R.~Alkofer and Chr.~Fischer for helpful discussions.


\begin{thebibliography}{sch90}
\setlength{\itemsep}{0pt}
\bibitem{bali95}{ G.~S.~Bali, K.~Schilling and C.~Schlichter,
   Phys. Rev.  {\bf D51} (1995) 5165; [hep-lat/9409005]. }
\bibitem{kap92}{ 
   D.~B.~Kaplan, Phys. Lett. {\bf B288} (1992) 342, 
   [hep-lat/9206013]. \\ 
   R.~Narayanan and H.~Neuberger,  Phys. Lett.  {\bf B302} (1993) 62, 
   [hep-lat/9212019]; Nucl. Phys. {\bf B412} (1994) 574, 
   [hep-lat/9307006]. \\ 
   P.~M.~Vranas, Nucl. Phys. Proc. Suppl. {\bf 94} (2001) 177, 
   [hep-lat/0011066]. } 
\bibitem{bar98}{ I.~M.~Barbour  [UKQCD Collaboration], Nucl. Phys. 
   {\bf A642} (1998) 251. } 
\bibitem{eng99}{ J.~Engels, O.~Kaczmarek, F.~Karsch and E.~Laermann,
   Nucl. Phys.{\bf B558} (1999) 307; [hep-lat/9903030]. \\ 
   K.~Langfeld and G.~Shin, Nucl. Phys. {\bf B572} (2000) 266, 
   [hep-lat/9907006]. } 
\bibitem{rob94}{ C.~D.~Roberts and A.~G.~Williams,
   Prog. Part. Nucl. Phys. {\bf 33} (1994) 477, [hep-ph/9403224]. } 
\bibitem{alk00}{ R.~Alkofer and L.~von Smekal,
   in press by Phys. Rep., [hep-ph/0007355]. } 
\bibitem{rob00}{ C.~D.~Roberts and S.~M.~Schmidt,
   Prog. Part. Nucl. Phys. {\bf 45S1} (2000) 1, 
   [nucl-th/0005064]. } 
\bibitem{bau98}{ L.~Baulieu and M.~Schaden,
   Int. J. Mod. Phys.{\bf A13} (1998) 985, [hep-th/9601039]. \\ 
   M.~Schaden and A.~Rozenberg, Phys. Rev. {\bf D57} (1998) 3670, 
   [hep-th/9706222]. } 
\bibitem{sme97}{ L.~von Smekal, R.~Alkofer and A.~Hauck,
   Phys. Rev. Lett. {\bf 79} (1997) 3591, [hep-ph/9705242]. \\ 
   L.~von Smekal, A.~Hauck and R.~Alkofer,
   Annals Phys. {\bf 267} (1998) 1, [hep-ph/9707327]. } 
\bibitem{cfi01}{ Chr.~Fischer, private communications. } 
\bibitem{atk98}{ D.~Atkinson and J.~C.~Bloch,
   Phys. Rev.{\bf D58} (1998) 094036, [hep-ph/9712459]. }
\bibitem{atk98b}{ D.~Atkinson and J.~C.~Bloch,
   Mod. Phys. Lett. {\bf A13} (1998) 1055, [hep-ph/9802239]. }
\bibitem{cuc97}{ A.~Cucchieri,
   Nucl.Phys. {\bf B508} (1997) 353, [hep-lat/9705005]. \\
   A.~Cucchieri and T.~Mendes,
   Nucl. Phys. Proc. Suppl. {\bf 53} (1997) 811, [hep-lat/9608051]. } 
\bibitem{bon01}{ F.~D.~Bonnet, P.~O.~Bowman, D.~B.~Leinweber, 
   A.~G.~Williams and J.~M.~Zanotti, hep-lat/0101013. \\ 
   F.~D.~Bonnet, P.~O.~Bowman, D.~B.~Leinweber and A.~G.~Williams,
   Phys. Rev.{\bf D62} (2000) 051501, [hep-lat/0002020]. } 
\bibitem{zwa94}{ D.~Zwanziger, Nucl. Phys. {\bf B412} (1994) 657. } 
\bibitem{deb97}{ L.~Del Debbio, M.~Faber, J.~Greensite and S.~Olejnik,
   Phys. Rev.{\bf D55} (1997) 2298, [hep-lat/9610005]. \\ 
   L.~Del Debbio, M.~Faber, J.~Giedt, J.~Greensite and S.~Olejnik,
   Phys. Rev.{\bf D58} (1998) 094501, [hep-lat/9801027]. } 
\bibitem{la98}{ K.~Langfeld, H.~Reinhardt and O.~Tennert,
   Phys. Lett. {\bf B419} (1998) 317, [hep-lat/9710068]. } 
\bibitem{la99}{ K.~Langfeld, O.~Tennert, M.~Engelhardt and H.~Reinhardt,
   Phys. Lett.{\bf B452} (1999) 301, [hep-lat/9805002]. \\ 
   M.~Engelhardt, K.~Langfeld, H.~Reinhardt and O.~Tennert,
   Phys. Rev. {\bf D61} (2000) 054504, [hep-lat/9904004].} 
\bibitem{for99}{ P.~de Forcrand and M.~D'Elia,
   Phys. Rev. Lett. {\bf 82} (1999) 4582, [hep-lat/9901020]. } 

\end{thebibliography}
\end{document}